# New 3D fine-grained scintillation detector for the T2K experiment

**S. Fedotov,**[a] **on behalf of the T2K ND280 Upgrade working group**

[a] *Institute for Nuclear Research of the Russian Academy of Sciences,
Prospekt 60-letiya Oktyabrya, 7a, Moscow, Russia*

*E-mail*: fedotov@inr.ru

ABSTRACT: The main goal of the T2K long-baseline neutrino experiment is the precise measurement of the parameters of neutrino oscillation and the search for CP-violation in the lepton sector. In 2017, the T2K collaboration launched the Near Detector Upgrade project. This upgrade aims to reduce the systematic errors of the oscillation parameters from 6-7% to 3-4%. In order to accomplish such goal, ND280 Upgrade proposes to implement new upstream trackers to improve the wide angle acceptance and low momentum threshold. One of the novel technologies introduced in ND280 Upgrade is super fine-grained (SuperFGD) scintillator target with 3D WLS fiber readout. The detector will consist of about 2 million (192×184×56) scintillation cubes of 1-cm size. Each cube has three orthogonal holes for WLS fibers to provide 3D optical readout. The total mass of the detector is about 2 tons and the number of the readout channels is about 60,000. This paper presents the current status of the development of this detector.

## Contents

## 1. Introduction

The main goal of the T2K long-baseline neutrino experiment [1] is the precise measurement of the parameters of neutrino oscillation and the search for CP-violation in the lepton sector. The main elements of experimental setup are the near detectors [2, 3] and the far detector Super-Kamiokande (SK), located 295 km away from the target. The near detector system, designed to monitor the neutrino beam parameters before oscillations, consists of two detectors: INGRID and ND280. The INGRID (Interactive Neutrino GRID) is located on the beam axis, and the ND280 is an off-axis detector, which is located at 2.5° to the beam axis.

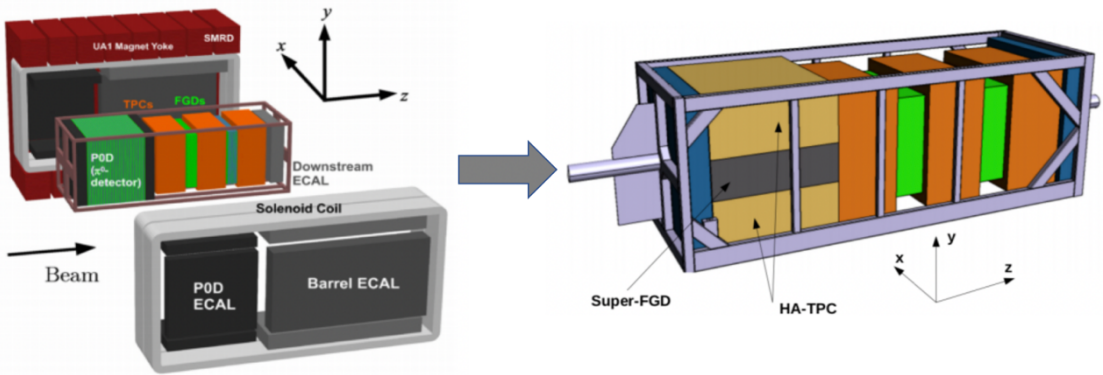

**Figure 1.** The off-axis detector. Left: before upgrade, Right: after upgrade.

The off-axis system ND280, shown in Fig. 1., consists of the tracker (2 fine-grained detectors FGD and 3 time projection chambers TPC) and other detectors (the $\pi^0$-detector P0D, electromagnetic calorimeter ECAL and side range muon detector SMRD), placed inside the UA1 magnet. In 2017, the T2K collaboration launched the Near Detector Upgrade project [4]. This upgrade aims to reduce the systematic errors of the oscillation parameters. In the upgraded system the P0D detector will be replaced by the segmented scintillator detector SuperFGD [5] and two horizontal TPCs. The need for an upgrade of the ND280 is caused by the following reasons: 1) the current tracker has an acceptance of the charged particles from neutrino interactions mostly in forward/backward direction, while the far detector SK has a 4π acceptance; 2) the sensitivity to CP-violation is limited by the systematics which could be reduced by precise measurements of the neutrino-interaction cross sections. The upgraded ND280 system will not only have a 4π-acceptance of the charged particles produced in the neutrino interactions inside the SuperFGD detector; it also will help to suppress the background from protons and pions and, probably, from neutrons, and reduce the systematic errors of the oscillation parameters from current 6-7% to 3-4% [6-7].

## 2. Design of the SuperFGD

Conception of the SuperFGD is shown in Fig. 2. The SuperFGD dimensions of an active volume are 192 (width) × 56 (height) × 184 (length) cubes. The full-size SuperFGD detector is comprised of 1,978,368 cubes and 56,384 readout channels with the active mass of ∼ 2 tons. The 1 mm diameter Kuraray Y-11 (200) Wavelength Shifting (WLS) fibers [8] are threaded through 1.5 mm diameter holes in each side of the cubes to readout the scintillation light from three views by Hamamatsu Multi-Pixel Photo-Counters (MPPCs). The $1\times1\times1$ cm$^3$ cube size is a natural granularity scale corresponding to the range of 200 MeV/c protons in plastic scintillator and strikes an acceptable balance between position resolution and number of readout channels. Signals from each cube will be read out via three orthogonal 1.0 mm Kuraray Y11 WLS fibers. The cubes are produced at Uniplast Co. from Vladimir (Russia) using injection molding. The scintillator composition is polystyrene doped with 1.5% of paraterphenyl (PTP) and 0.01% of 1.4-bis benzene (POPOP). Once each cube was produced, a reflective layer was etched onto each of the cubes surfaces with a chemical agent, resulting in the formation of a 50 – 80 μm-thick white polystyrene micropore deposit. Each cube was placed on a jig designed to hold it in place during the drilling of the three orthogonal through holes of 1.5 mm diameter. A Computer Numerical Controlled (CNC) milling machine was used to drill the holes. The cubes have a side width of 10.026 mm right after the molding. Reflector covering increases the size of the cubes to 10.167 mm ± 30 μm. The fluctuations of the hole positions relative to the cube sides is less than 50 μm (rms).

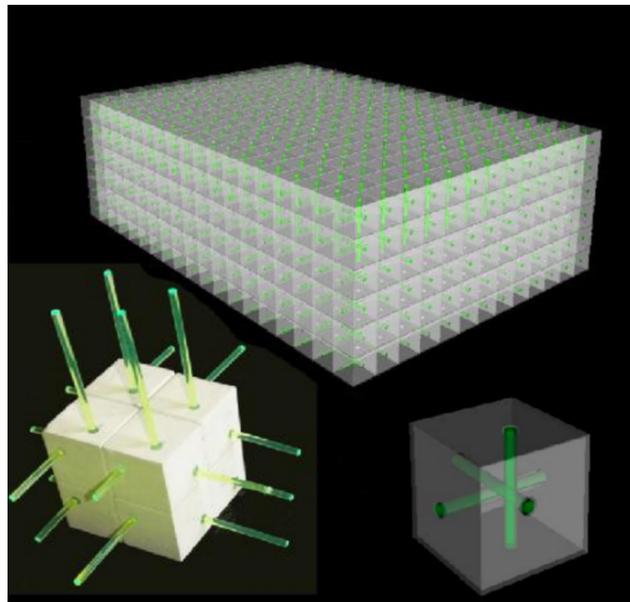

**Figure 2.** The design of the SuperFGD. The group of 8 cubes illustrates the readout method. A single fiber is going through a row of cubes. One end of the fiber is viewed by a photosensor, another end is covered by a reflector. Each cube is viewed by three orthogonal fibers.

## 3. Beam tests

A 5×5×5 cube array was tested at CERN in autumn 2017 (results reported in [9, 10]). The signal is read out by Kuraray Y11 WLS s-type fibers 1 mm in diameter and 1.3 m long, and also by Hamamatsu MPPCs 12571-025C with the 1×1 mm$^2$ active area, which contains 1600 pixels. A ~5 GHz digitizer was used to read out 12 fibers out of 75 mounted ones to measure the parameters of the prototype. Typical light yield (L.Y.) from a single cube per a fiber was measured to be 40 p.e./MIP, two readout fibers produced L.Y. of about 80 p.e./MIP. Time resolution per a single fiber was 0.95 ns, two fibers provided the resolution of 0.65-0.71 ns. Optical crosstalk through a side of a cube was found to be around 4%. The first promising results motivated the assembly and testing of a larger prototype, a 24×8×48 cube array (x-y-z detector axes) of 9,216 cubes instrumented with 1,728 readout channels. This prototype was tested extensively at the T9 beamline of the Proton Synchrotron (PS) accelerator at CERN in the summer of 2018. The length of the fibers was equal to the length of the prototype sides. Fig.3 shows the examples of the tracks recorded in this beam test. The obtained data have been analyzed and the publication is currently in preparation.

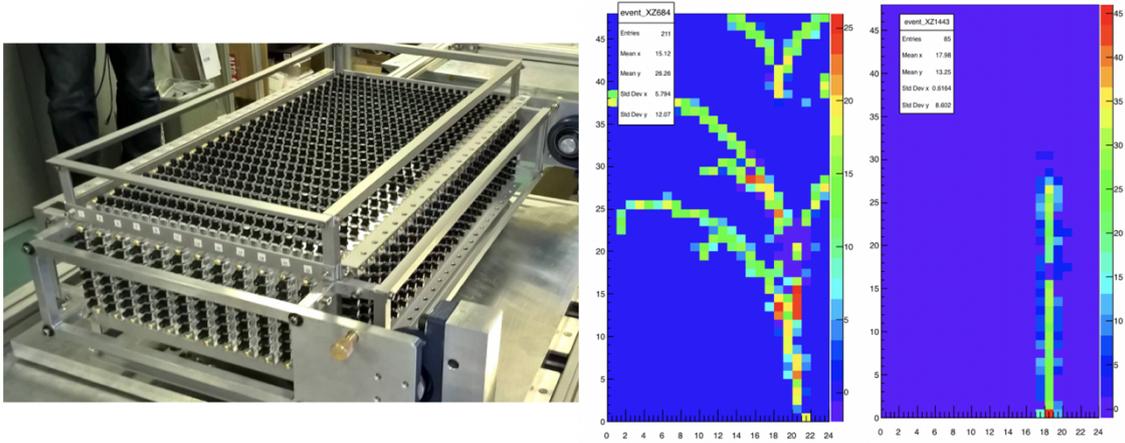

**Figure 3.** Left: the second prototype. Right: examples of the events from the second prototype: multi track event and stopped proton track.

## 4. Kuraray Y11 WLS-fibers tests

Kuraray Y11 WLS fibers of 1 mm diameter will be used to readout scintillating signals in the SuperFGD. We have studied the attenuation in the fibers with lengths equal to the side dimensions of the detector: 60 cm and 2 m. One end of the fiber was viewed by the Hamamatsu MPPC S13081-050CS, the other one was either polished, polished and covered with reflective paint or black paint. The fibers were illuminated by a scintillator cube excited by an UV diode (380 nm). The results are presented in Fig. 4. The double-exponential function was used for fitting the attenuation points. The fibers, the end of which was polished and covered with reflective paint, showed the best result as expected. Their L.Y. was 30% higher for the 60-cm long samples and 40% higher for 2-m long samples relative to untreated fibers.

    The consistent results were obtained in the tests with cosmic-ray muons though with lower accuracy. For cosmic tests the «small» prototype of 5×5×5 cubes was used as a light source for the WLS fibers. The tested fiber passed through the central cubes. The cubes located above and below the measured ones were used as trigger counters.

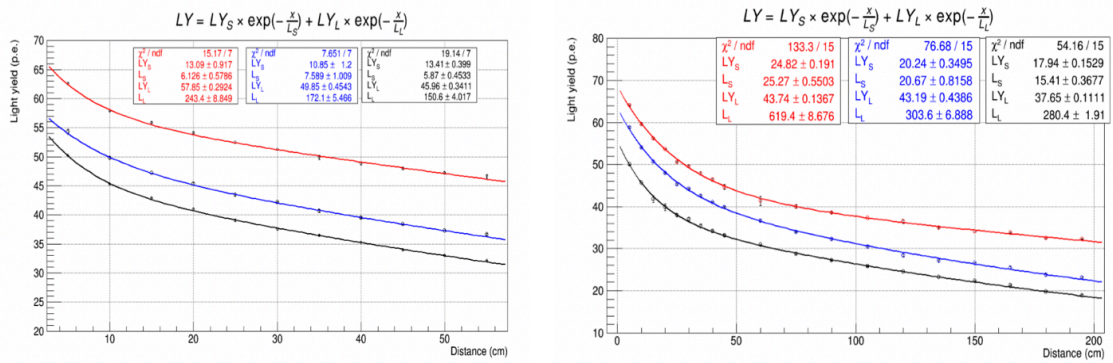

**Figure 4.** The dependence of L.Y. on the distances to the MPPC for the Kuraray Y11 fibers (60 cm and 2 m long.) The red line: the end of a fiber which was polished and covered with reflective paint; the blue line: the end of a fiber which was just polished; the black line: the end of a fiber which was polished and covered with black paint.

## 5. Neutron detection

A new method for an improved anti-neutrino energy reconstruction presented in detail in the article [7]. This method is based on the precise measurements of the outgoing neutron kinetic energy. A neutron detected through the identification of a proton, which was born at a secondary neutron interaction (Fig. 5). An anti-neutrino interacts with nucleus to produce a muon and a neutron. After some time, this neutron interacts ejecting a proton. The neutron energy is reconstructed through measurements of the proton energy and the time between two events. In this method, the neutron detection efficiency for the SuperFGD detector is relatively high (Fig. 6). It is expected that this will improve the anti-neutrino energy resolution to 7% relative to 15%, which expected using tradition neutrino energy reconstruction techniques.

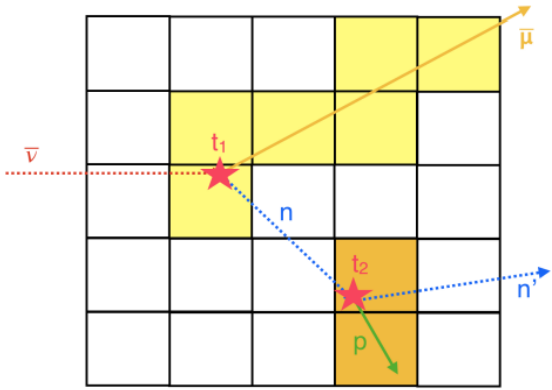
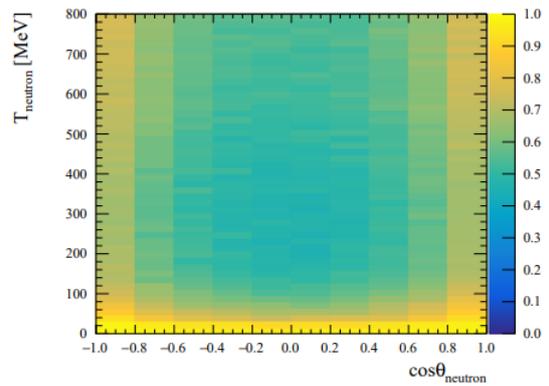

**Figure 5.** Example of the neutron detection through the observation of secondary proton.

**Figure 6.** The detection efficiency (Monte Carlo simulation).

## 6. Assembly

The method to assemble many thousands of the cubes into 3D structure is based on using a fishing line of 1.3 mm calibrated diameter, which is threaded through the cubes. The cubes are assembled into array using the fishing lines, which forms the 3D skeleton structure with specified position for each cube. After that the fishing lines are removed and the WLS fibers are inserted in place one by one. A linear chain of cubes on a fishing line is the basic element for the assembling process. The linear chains are sewn together into 2D flat planes (Fig. 7), also using fishing lines. During the final stage, the cube planes are merged vertically with needles and fishing lines into a 3D body. Mock-ups and prototypes of differing size were assembled to test the fishing line method. One of the prototypes consists of 5 full size planes of 192×184 cubes. Another real size prototype consists of 56 narrow planes of 192×15 cubes each. The tests demonstrated that the fibers can be smoothly inserted through the cube rows over a 2 m length. All tested fishing lines were easily replaced without any problems. In total, more than 100 channels were tested. After installation of WLS fibers, the MPPC boards and readout electronics will be mounted.

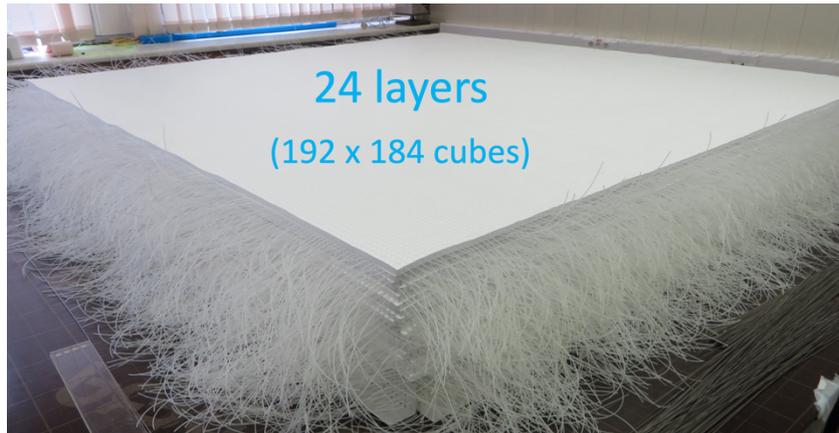

**Figure 7.** 24 planes of cubes assembled on the fishing line. The fishing lines on the Z-axis are not installed.

The quality check of all the cubes is performed before the assembling. The first step is to assure the correct geometry of each cube and accurate positions of 3 holes in a cube. To reach this objective we form 14×14 cube array on the 1.4 mm diameter needles. All the cubes have to slide smoothly along the needles in x-y directions inside the array. Quality of scintillator is also under permanent control. We randomly select cubes from different shipments and test them with cosmic-ray muons. Tested configuration is the following: three cubes per one Kuraray Y11 fiber. 8 fibers with 24 cubes are measured simultaneously. One end of the fibers is viewed by Hamamatsu MPPC S13081-050C with 1.3×1.3 mm$^2$ sensitive area. The other end of the fiber is polished. About 1200 cubes were tested during the period of observation. L.Y. was found to be stable, no defective cubes were found. The average L.Y. was ~ 36 p.e./MIP.

## 7. Conclusion

The new near detector (SuperFGD) for the T2K experiment is in the active stage of the development. Currently (the spring of 2020), a half of the detector (1 million cubes) on the fishing lines is completed in the form of assembled cube planes. The replacing the fishing lines with WLS fibers was tested on several full-size prototypes. The main parameters of the future detector were measured during the beam tests at CERN: typical L.Y. – 40 p.e./MIP/fiber; time resolution – 0.95 ns/fiber; crosstalk – 4 % per one side of a cube. The SuperFGD is planned to be installed into the UA1 magnet in October 2021.

**Acknowledgements**

This work is partly supported by the JSPS-RFBR grant #20-52-50010 and by the RSF grant #19-12-00325. The author is grateful to the ND280 Upgrade Working group for their cooperation and detailed discussions regarding the SuperFGD. Special thanks are due to Yu. Kudenko, M. Khabibullin and O. Mineev for discussions and useful comments.